\documentclass[useAMS,usenatbib]{mn2e}

\usepackage{url,times,graphicx,amsmath,amsfonts,amssymb,aas_macros,color,epsfig,epstopdf}

\newcommand{\Rmax}{$R_{\rm max}$}
\newcommand{\Vmax}{$V_{\rm max}$}
\newcommand{\nPS}{$n_{\rm P\&S}$}
\newcommand{\nEin}{$n_{\rm E}$}

\newcommand{\hMpc}{{\ifmmode{h^{-1}{\rm Mpc}}\else{$h^{-1}$Mpc}\fi}}
\newcommand{\hkpc}{{\ifmmode{h^{-1}{\rm kpc}}\else{$h^{-1}$kpc}\fi}}
\newcommand{\hMsun}{{\ifmmode{h^{-1}{\rm {M_{\odot}}}}\else{$h^{-1}{\rm{M_{\odot}}}$}\fi}}
\newcommand{\ltsima}{$\; \buildrel < \over \sim \;$}
\newcommand{\gtsima}{$\; \buildrel > \over \sim \;$}
\newcommand{\lsim}{\lower.5ex\hbox{\ltsima}}
\newcommand{\gsim}{\lower.5ex\hbox{\gtsima}}
\def\LCDM{$\Lambda$CDM}

\def\lesssim{\mathrel{\hbox{\rlap{\hbox{\lower4pt\hbox{$\sim$}}}\hbox{$<$}}}}
\def\gtrsim{\mathrel{\hbox{\rlap{\hbox{\lower4pt\hbox{$\sim$}}}\hbox{$>$}}}}

\newcommand{\Tab}[1]{Table~\ref{#1}}
\newcommand{\Sec}[1]{Section~\ref{#1}}

\newcommand{\Fig}[1]{Fig.~\ref{#1}}
\newcommand{\beq}{\begin{equation}}
\newcommand{\eeq}{\end{equation}}
\def\beqa{\begin{eqnarray}}
\def\eeqa{\end{eqnarray}}
\def\hMpc{$h^{-1}\,{\rm Mpc}$}
\def\hkpc{$h^{-1}\,{\rm kpc}$}
\def\LCDM{\ensuremath{\Lambda}CDM}

\def\Vmax{$V_{\rm max}$}
\def\Rmax{$R_{\rm max}$}
\def\head{
 \vbox to 0pt{\vss
                   \hbox to 0pt{\hskip 440pt\rm LA-UR-10-07069\hss}
                  \vskip 25pt}}
\title[Density profile of subhaloes in SPH simulation]
{Size matters: the non-universal density profile of subhaloes in SPH simulations and implications for the Milky Way's dSphs}
\author[Di Cintio et. al]
       {Arianna Di Cintio$^{1,2,3}$\thanks{E-mail: arianna.dicintio@uam.es}, Alexander Knebe$^{1}$, Noam I. Libeskind$^3$, Chris Brook$^1$, \newauthor  Gustavo Yepes$^1$, Stefan Gottl\"ober$^3$, Yehuda Hoffman$^4$\\
$^{1}$Departamento de F\'isica Te\'orica, M\'odulo C-15, Facultad de Ciencias, Universidad Aut\'onoma de Madrid, 28049 Cantoblanco, Madrid, Spain\\
$^2$Physics Department "G. Marconi", Universita' di Roma "Sapienza", Ple Aldo Moro 2, 00185 Rome, Italy\\
$^3${Leibniz-Institut f\"{u}r Astrophysik} Potsdam, An der Sternwarte 16, D-14482 Potsdam, Germany\\
$^4$Racah Institute of Physics, The Hebrew University of Jerusalem, Givat Ram, Israel
}

\begin{document}

\date{Accepted XXXX . Received XXXX; in original form XXXX}

\pagerange{\pageref{firstpage}--\pageref{lastpage}} \pubyear{2010}

\maketitle

\label{firstpage}

%\clearpage

%%%%%%%%%%%%%%%%%%%%%%%%%%%%%%%%%%%%%%%%%%%%%%%%%%%
\begin{abstract}
We use dark matter only and full hydrodynamical Constrained Local UniversE Simulations (CLUES) of the formation of the Local Group     
to study the density profile of subhaloes of the simulated Milky Way and Andromeda galaxies. We show that the Einasto model provides the best description of the subhaloes' density profile, as opposed to the more commonly used NFW profile or any generalisation of it. We further find that the Einasto shape parameter \nEin\ is strongly correlated with the total subhalo mass, pointing towards the notion of a non-universality of the subhaloes' density profile. We observe that the effect of mass loss due to tidal stripping, in both the dark matter only and the hydrodynamical run, is the reduction of the shape parameter \nEin\ between the infall and the present time. Assuming now that the dSphs of our Galaxy follow the Einasto profile and using the maximum and minimum values of \nEin\  from our hydrodynamical simulation as a gauge, we can improve the observational constraints on the \Rmax-\Vmax\ pairs obtained for the brightest satellite galaxies of the Milky Way. 
When considering only the subhaloes with $-13.2\lesssim M_V\lesssim-8.8$, i.e. the range of luminosity of the classical dwarfs, we find that all our simulated objects are consistent with the observed dSphs if their haloes follow the Einasto model with $1.6\lesssim n_{\rm E} \lesssim5.3$. The numerically motivated Einasto profile for the observed dSphs will alleviate the recently presented "massive failures" problem.
\end{abstract}

%%%%%%%%%%%%%%%%%%%%%%%%%%%%%%%%%%%%%%%%%%%%%%%%%%%
\noindent
\begin{keywords}
 methods: numerical - $N$-body simulations -- galaxies: formation - haloes - Local Group
 \end{keywords}

%%%%%%%%%%%%%%%%%%%%%%%%%%%%%%%%%%%%%%%%%%%%%%%%%%%
\section{Introduction} \label{sec:introduction}
%%%%%%%%%%%%%%%%%%%%%%%%%%%%%%%%%%%%%%%%%%%%%%%%%%%
While the predictions of the current $\Lambda$ Cold Dark Matter (\LCDM) model have been widely confirmed at cosmological scales, there are still a number of discrepancies between theory and observations at galactic and subgalactic scales: one example is  the well-known "missing satellite problem", first pointed out by \citet{Klypin99} and \citet{Moore99}. The high number of substructures resolved within the virial radius of galaxy-type objects in high resolution cosmological simulation mismatches the number of observed satellite galaxies of our Milky Way (MW) and nearby galaxies. To alleviate the problem one must invoke some mechanisms, such as early reionization of the intergalactic medium and supernovae feedback \citep{Bullock00,Somerville02,Benson02}, to suppress galaxy formation below a certain scale.

However, there is an inconsistency not only with the number, but also about the kinematics of the observed MW's dwarf spheroidals (dSphs) when compared to the velocity profiles of the most massive subhaloes found in dark matter simulations \citep{Boylan11}. Assuming that these subhaloes follow a \citet[][NFW herafter]{Navarro96} profile, they have been found to be too dense to host the MW's bright satellites. This is directly related to the findings of \citet{Bovill11a,Bovill11b}, whose simulations showed an overabundance of bright dwarf satellites ($L_V > 10^4L_{sun}$) with respect to the MW's dSphs.

A number of studies tried to reconcile simulations with observations.

The possibility that the MW is a statistical outlier has been ruled out by \citet{Strigari12}, who used data from the Sloan Digital Sky Survey to show that, down to the scale of Sagittarius dwarf, our Galaxy is not anomalous in its number of classical satellites.
Further, the analysis of \citet{Boylan12}, independent from the choice of actual density profile of the simulated subhaloes, demonstrates that supernova feedback is unlikely to be an explanation for the low inferred densities of dSphs, due to their small stellar masses.
Different hypothesis for the nature of dark matter can naturally lead to the formation of less concentrated subhaloes in a warm dark matter scenario \citep{Lovell11} or in simulations of self-interacting dark matter models \citep{Vogelsberger12}, providing an interesting alternative to the \LCDM\ model.
Moreover, the discrepancy between observed and simulated satellite galaxies may reflect the fact that the MW is less massive than is commonly
thought: a total mass between $8\cdot10^{11}\lesssim M/M_{\odot}\lesssim10^{12}$ has been argued in \citet{Vera-Ciro12} and \citet{Wang12}.
However, lowering the mass of the MW still do not explain why its dSphs (as well as many isolated dwarf galaxies), seem to live in haloes whose mass is smaller than the current expectation from abundance matching models \citep{Ferrero11}.

The inclusion of baryons in simulations has also been explored, and it has been found either to have negligible effects on the dark matter density of subhaloes \citep{Parry11} or to have a twofold effect on their density profile \citep{DiCintio11}.
In fact, recognizing that at galactic scales baryonic processes are expected to play a crucial role, it has been investigated in \citet{DiCintio11} the effect of the inclusion of baryons in SPH simulation within the CLUES project.\footnote{http://www.clues-project.org} These simulations are designed and constrained, respectively, to reproduce as closely as possible the actual observed Local Group with its two prime galaxies MW and Andromeda (hereafter also referred to as M31) and hence serve as an ideal testbed for investigating the dynamics and kinematics of the satellite populations of the real MW and M31. In this previous study it has been found that, while in some cases the baryons are able to lower the central density of subhaloes, through mechanisms such as gas outflows driven by star formation and supernovae \citep{Navarro96b,Governato12}, there are still substructures whose density is increased, as expected from the adiabatic contraction model of \citet{Blumenthal86}.

The underlying assumption in many previous works is that the satellite galaxies of the MW are embedded in subhaloes whose mass profile is described by the NFW model: it is still a matter of debate, however, if this profile is the best choice in modeling the dSphs' density.    

On one hand, \citet{Walker11} constructed a method for measuring the slope of the mass profiles within dSphs directly from stellar spectroscopic data, independently from any dark matter halo model and velocity anisotropy of the stellar tracers, and showed evidence for the profile of the Fornax and Sculptor dSphs to be consistent with cores of constant density within the central few-hundred parsecs of each galaxy, thus ruling out a cuspy profile such as the NFW one. 
On the other hand, \citet{Wolf12} used a Jeans analysis to show that, even in the limiting case of an isotropic velocity dispersion, not
all of the dwarfs prefer to live in halos that have constant density cores. It must be noticed, however, that while the \citet{Walker11} method is insensitive to orbital anisotropy and underlying halo potential, the \citet{Wolf12} results are dependent from these yet unknown quantities.

Regarding simulations, \citet{DiCintio11} pointed out that the NFW profile may not be appropriate to describe the subhaloes' density. While computing the subhaloes' circular velocities \citet{Boylan12} used an Einasto profile to model the density distribution of subhaloes at small radii, in order to correct for the effects of the force softening, and the raw particle data at higher radii.

The three-parameter Einasto profile \citep{Einasto65}, indeed, has been found to more accurately describe the halo density in dark matter only simulations \citep[e.g.][]{Navarro04,Merritt06I,Prada06,Gao08,Hayashi08,Navarro10,Ludlow11,Reed11}, reducing the residuals of the fits by $20\%$ with respect to the corresponding NFW models. 

In this \textit{Paper} we mainly focus on hydrodynamical simulations and, after a brief explanation, in \Sec{sec:simulation}, of the CLUES project simulations, we study the mass profile of substructures within the two main haloes of the simulated Local Group, formally calling them Milky Way and M31. In \Sec{sec:density} we focus on the quality of several analytical models in describing the density profile of galactic subhaloes showing that, also in hydrodynamical simulations, the Einasto profile provides the best description. In \Sec{sec:n_dependence} we then show that the corresponding profile shape parameter \nEin\ scales with the virial mass of the subhalo. We finally discuss the implications for the mismatch between the kinematics of the observed MW's dSphs and the simulated substructures in \Sec{sec:observation}, before concluding in \Sec{sec:conclusion}.

%%%%%%%%%%%%%%%%%%%%%%%%%%%%%%%%%%%%%%%%%%%%%%%%%%%
\section{The Simulations} \label{sec:simulation}
%%%%%%%%%%%%%%%%%%%%%%%%%%%%%%%%%%%%%%%%%%%%%%%%%%%
Our simulations form part of the aforementioend CLUES project and are based upon a WMAP3 cosmology. These constrained simulations of the Local Universe have already been presented and extensively used for other investigations and we refer the reader to those articles for more details \citep[e.g.][]{Gottloeber10,Libeskind10,Libeskind10infall,Knebe10a,Knebe11a,DiCintio11}. We therefore only repeat here the most basic informations. The simulations assume a WMAP3 cosmology \citep{Spergel07}, i.e. $\Omega_m = 0.24$, $\Omega_{b} = 0.042$, $\Omega_{\Lambda} = 0.76$ and $h=0.73$, a normalization of $\sigma_8 = 0.75$ and a slope of the power spectrum of $n=0.95$. The treePM-SPH code \texttt{GADGET2} \citep{Springel05} has been applied to simulate the evolution of a cosmological box with side length of $L_{\rm box}=64 h^{-1} \rm Mpc$ in which the formation a Local Group has been enforced by constraints on the initial conditions. There are two runs available, one with dark matter only (DM run) and one hydrodynamical (labelled SPH run) in which we additionally follow the feedback and star formation rules of \cite{Springel03}, as well as a uniform but evolving ultra-violet cosmic background \citep{Haardt96}. The runs feature a mass resolution of $m_{\rm  DM}=2.1\times 10^{5}$\hMsun\ for the dark matter particles ($m_{\rm  DM}=2.54\times 10^{5}$\hMsun\  in the DM only run), $m_{\rm  gas}=4.42\times 10^4$\hMsun\ for the gas particles and $m_{\rm  star}=2.21\times 10^4$\hMsun\ for the star particles. The gravitational softening length is $\epsilon=0.1 h^{-1}kpc=137pc$, in both the DM only and the SPH run. 

The stellar population synthesis model STARDUST \citep[see][and references therein for a detailed description]{Devriendt99} has been used to derive luminosities from the stars formed in our simulation. This model computes the spectral energy distribution from the far-UV to the radio, for an instantaneous starburst of a given mass, age and metalicity. The stellar contribution to the total flux is calculated assuming a Kennicutt initial mass function \citep{Kennicutt98}. 

The \texttt{AHF} halo finder\footnote{http://popia.ft.uam.es/AMIGA} has been used to identify all (sub-)haloes in our simulation. Note that \texttt{AHF} automatically (and essentially parameter-free) finds haloes, sub-haloes, sub-subhaloes, etc. All the subhaloes used in this particular study are selected in order to be more massive than $M_{\rm sub}\geqslant2\times10^8h^{-1}M_{\odot}$, with a peak in the velocity curve $V_{\rm max}\gtrsim10$km/s, and to lie within $300$ kpc from each host's center, the latter being either the MW or M31. The masses of the SPH hosts, defined as the masses within a sphere containing $\Delta_{vir}\simeq390$ times the cosmic mean matter density, are $M_{\rm MW}=4.0\times10^{11}h^{-1}M_\odot$ and $M_{\rm M31}=5.47\times10^{11}h^{-1}M_\odot$. When stacking the data from the two hosts together, we found a total of 56 SPH and 66 DM subhaloes in this WMAP3 simulation. Note that our selection criterion assures that within each host a subhalo contains a minimum of 1000 particles.

%%%%%%%%%%%%%%%%%%%%%%%%%%%%%%%%%%%%%%%%%%%%%%%%%%%%%
\section{The density profile of SPH and DM subhaloes}\label{sec:density}
%%%%%%%%%%%%%%%%%%%%%%%%%%%%%%%%%%%%%%%%%%%%%%%%%%%%%
%%%%%%%%%%%
\begin{table*}
 \caption{Quality of the fits for several density profile models. The results for SPH and DM subhaloes are listed, together with the mean value of the shape parameter $n$, the inner slope $\gamma$ and the outer slope $\beta$, respectively.}
\begin{center}
\begin{tabular}{lcccc}
\hline
\hline

& \multicolumn{2}{c}{SPH} & \multicolumn{2}{c}{DM} \\

\hline
Profile  & $\overline{\Delta^2}$ & shape & $\overline{\Delta^2}$ & shape \\
\hline
NFW                     & $0.043$ & $\gamma=1.00$                          & 0.042    &   $\gamma=1.00$                                                      \\
M99                      & $0.030$ & $\gamma=1.50$                            &    0.032  &  $\gamma=1.50$                                                   \\
$(1,3,\gamma)$ & $0.014$ & $\overline{\gamma}=1.98$            &     0.015  &  $\overline{\gamma}= 1.53  $                           \\
$(1,\beta,1)$       & $0.014$ & $\overline{\beta}=3.80$                   &     0.013 &  $\overline{\beta}= 4.40  $                           \\                                                                                        
P\&S                     & $0.013$ & $\overline{n}_{\rm PS} =3.35$        &    0.013 &  $\overline{n}_{\rm PS} =3.15$                         \\
Einasto                & $\textbf{0.011}$ & $\overline{n}_{\rm E} =4.80$   & \textbf {0.012} & $\overline{n}_{\rm E} =  3.79 $           \\
\hline
\hline
\end{tabular}
\end{center}
\label{tab:profiles}
\end{table*}	
%%%%%%%%%%%

\subsection{Theoretical Models}
%%%%%%%%%%%%%%%%%%%%%%%%%%%%%%%%%%%%%%%%%%%%%%%%%%%%%
While it is widely accepted that a \citet[][NFW]{Navarro96} profile provides a good description of DM haloes, it has been already pointed out in \citet{DiCintio11} that this universal profile may not be the best choice when used to fit \textit{sub}-halo densities. We will thus study different profiles and apply them to our simulated substructures, with particular emphasis on the density profile of subhaloes in hydrodynamical simulations.

\paragraph*{Double-power law profiles}
A generalisation of the NFW profile is the so-called $(\alpha,\beta,\gamma)$, or double power-law model:

\beq
\rho_{\alpha,\beta,\gamma}(r)=\frac{\rho_s}{\left(\frac{r}{r_s}\right)^{\gamma}\left[1 + \left(\frac{r}{r_s}\right)^{\alpha}\right] ^{(\beta-\gamma)/\alpha}} 
\eeq

\noindent
where $r_s$ is the scale radius and $\rho_s$ the scale density, characteristic of each halo and related to its formation time and mass \citep[e.g.][]{Prada12,Munoz11,Maccio07,Bullock01}. It is a five-parameter model in which the inner and outer region have logarithmic slopes $-\gamma$ and $-\beta$, respectively, and the $\alpha$ parameter regulates the sharpness of the transition.
The choice $(\alpha,\beta,\gamma)=(1,3,1)$ provides the NFW profile, while $(\alpha,\beta,\gamma)=(1.5,3,1.5)$ gives the model presented in \citet[][M99 hereafter]{Moore99}. Besides of the NFW and M99 profiles we will also investigate the case of leaving the central slope as a free parameter, i.e. a $(1,3,\gamma)$ model.

\paragraph*{Einasto profile}
In addition to these double-power law profiles we test the Einasto profile \citep{Einasto65}, identical in functional form to the $2D$ S\'ersic model \citep{Sersic63, Sersic68}, but used instead to fit a spacial mass density:

\beq
\rho_{\rm E}(r)= \frac{\rho_{-2}}{e^{2n\left[\left(\frac{r}{r_{-2}}\right)^\frac{1}{n}-1\right]}}\label{einasto}
\eeq

\noindent
Here $r_{-2}$ is the radius where the logarithmic slope of the density profile equals -2 and $n$, also referred to as \nEin, is a parameter that describes the shape of the density profile. $r_{-2}$ is equivalent to the scale radius $r_s$ of a NFW profile, and the density $\rho_{-2}=\rho(r_{-2})$ is related to the NFW one through $\rho_{-2}=\rho_s/4$.  This profile gives a finite total mass and its logarithmic slope decreases inwards more gradually than a NFW or M99 profile. When \nEin\ is large, the inner profile is steep and the outer profile is shallow. Typical values of \nEin\ found in dark matter only simulations for haloes more massive than $10^{10}M_\odot$ are $4\lesssim n\lesssim7$.

\paragraph*{Prugniel--Simien profile}
Finally, following the study of \citet{Merritt06I}, we use the analytical approximation of the deprojected S\'ersic law, given in \citet[][P$\&$S hereafter]{Prugniel97}:

\beq
\rho_{\rm P\&S}(r)= \frac{\rho_{-2}}{ \displaystyle \left(\frac{r}{r_{-2}}\right)^p    e^{n(2-p)\left[\left(\frac{r}{r_{-2}}\right)^\frac{1}{n}-1\right]}}\label{prugniel}
\eeq

\noindent
where $n$, or \nPS\ where appropriate, is again a parameter describing the curvature of the density profile and the quantity $p$ is a function of \nPS\ chosen to maximize the agreement between the P$\&$S model and the S\'ersic law.  A good choice for $p$, when $0.6\lesssim n \lesssim10$, is $p=1.0 -0.6097/n + 0.05463/n^2$ \citep{Lima99}, used in our fitting routine. We must highlight that the shape parameter \nEin\ of the Einasto profile is not the same as \nPS\ of the P$\&$S model, although they follow the same functional form. The Einasto profile, the P$\&$S one and the modified NFW profile $(1,3,\gamma)$ are all 3-parameters models.

%%%%%%%%%%%%%%%%%%%
\begin{figure*}\begin{center}
  \includegraphics[width=18cm]{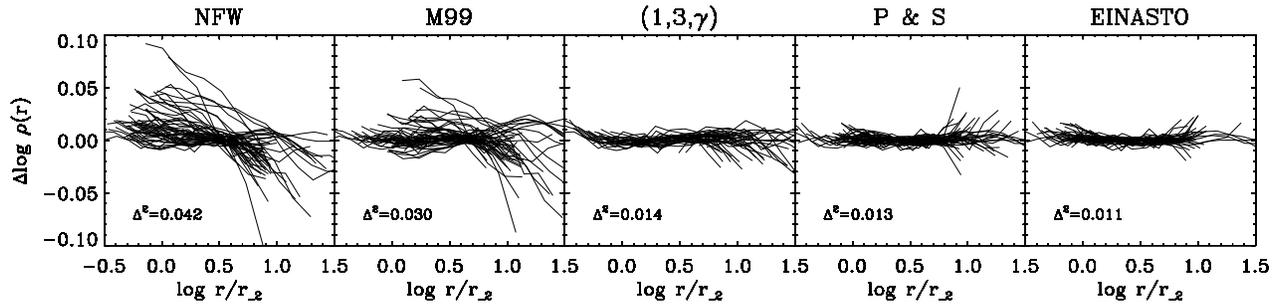}
\caption{Residuals of the density profiles of all SPH subhaloes for each of the fitted models. The mean goodness-of-fit  $\overline{\Delta^2}$ is indicated, providing the Einasto model to be the best one. The radial dependence of the residuals is the same for the DM only run, thus not shown here.}\label{fig:residuals}
\end{center}\end{figure*}
%%%%%%%%%%%%%%%%%%%

\begin{figure*}\begin{center}
$\begin{array}{cc}
   \includegraphics[width=3.3in]{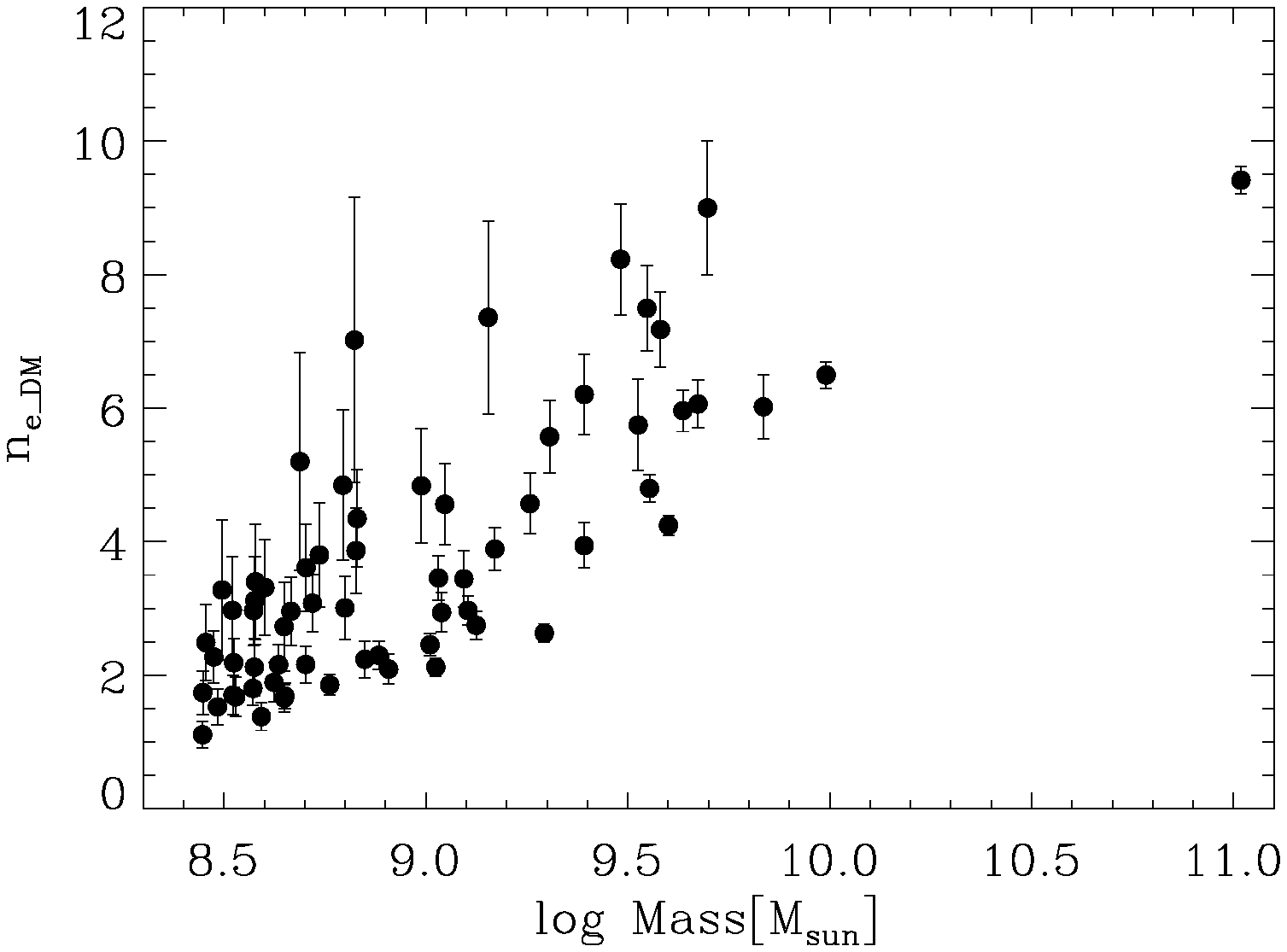}
  \includegraphics[width=3.3in]{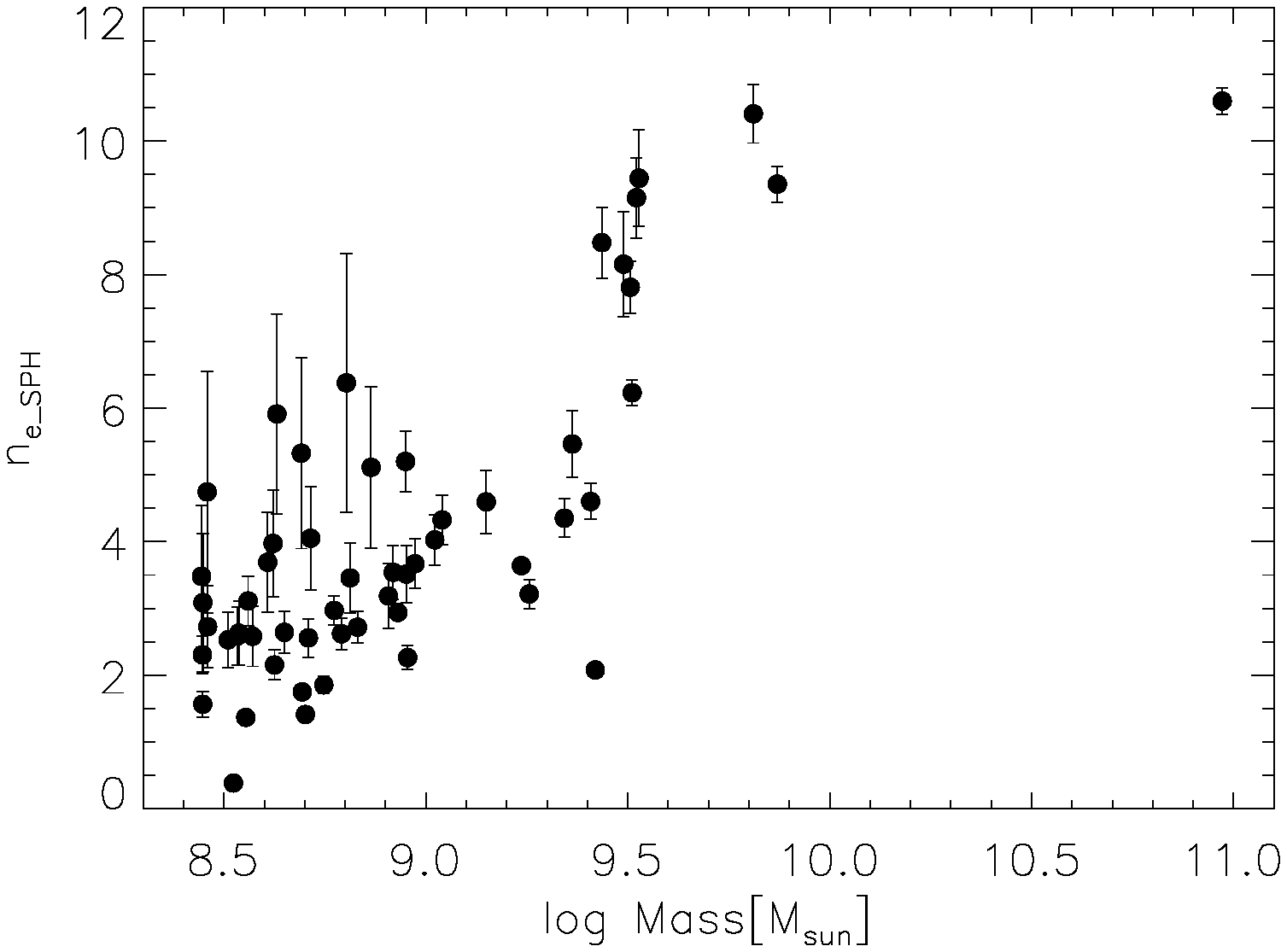}
\end{array}$
\caption{Correlation of the shape parameter \nEin\  with the subhaloes' masses in the DM only simulations, left panel, and SPH simulation, right panel. The error bars associated to the shape parameter are indicated for \nEin\ , as provided by the fitting routine. The $1\sigma$ statistical error committed in the evaluation of \nEin\  is, on average, the $15\%$ of the \nEin\ itself. The evident correlation between \nEin\ and the subhaloes' masses in the DM only as well as in the SPH run is attributed to the dynamical effects of tidal stripping. The additions of baryons can ulteriorly modify the density profile, specifically in its inner part.}
\label{fig:nEin_mass}
\end{center}
\end{figure*}

\subsection{Application to Subhaloes}
%%%%%%%%%%%%%%%%%%%%%%%%%%%%%%%%%%%%%%%%%%%%%%%%%%%%%
We now apply all the above models to fit our subhaloes' density profiles.\footnote{We use the IDL routine MPFIT}
The density profiles are given in radial bins logarithmically spaced from the inner radius compliant with the convergence criterion of \citet{Power03} out to the subhaloes' edge, defined as in \citet{Knollmann09}. The number of bins varies from 7 for the least massive objects to 16 for the most massive ones; by this we assure to minimize the Poissonian noise always having at least 150 particles per bin, and a minimum of 1000 particles in total in each subhalo. 
We verified that the convergence criterion as defined in \citet{Power03} is suitable also for subhaloes, and thus fully applicable to our simulation. Specifically, we used a lower resolution, $2048^3$ particles DM-only run with three times higher softening length $\epsilon=411$~pc, to show that the density profile of subhaloes converges for $\sim4.8\epsilon$. This value is always equal or less than the radius found using the \citet{Power03} criterion: all our trusted radii are thus fully converged according to the most conservative criterion possible and are not affected by two-body relaxation effects.

We define the goodness-of-fit as

\beq
\Delta^2=\frac{1}{N_{\rm bins}}\sum_{k=1}^{N_{\rm bins}}(\rm log_{10}\rho_{sim,k} - log_{10}\rho_{fit,k})^2,
\eeq

\noindent
whose average value over the total number of subhaloes $\overline{\Delta^2}$ gives an indication of the fit performance.

The results are presented in \Tab{tab:profiles} for a WMAP3 cosmology, where we list the quality of fit values $\overline{\Delta^2}$ alongside the mean value of the shape parameter \nEin, \nPS\ or the inner (outer) slope $\gamma$ ($\beta$) in the case of considering the double power-law models. 
A first consideration regards the differences between the DM only and the SPH runs: we observe that the mean shape parameters in the SPH run are systematically higher than in the DM counterpart, which implies a more cuspy central slope, indicating that the net effect of the inclusion of baryons is a steepening of the subhaloes' density. Our results appear to be in agreement with the prescription of an adiabatic contraction model \citep{Blumenthal86}, as shown already in \citet{DiCintio11} for the most massive and most luminous subhaloes.
We must remark that we are listing the average $n$ over the total set of subhaloes: there are cases, as discussed in \citet{DiCintio11}, in which the SPH subhaloes with the lowest baryon fraction, instead, undergo an expansion, therefore lowering their $n$. 
An higher shape parameter in the SPH run is also indicative of a less steep outer profile with respect to the DM only run, which means that tidal stripping effects are stronger on the DM only substructure, as reported in \citet{Libeskind10} (see discussion below).

We also notice that the mean shape parameter of the P\&S profile, \nPS, is lower than the corresponding Einasto parameter, \nEin, in the same run: this is expected, and found also in \citet{Merritt05}.
A few words on the steep central slopes $\gamma$ found for the $(1,3,\gamma)$ model: this model imposes the outer slope to be equal to $3$, which is not the case for subhaloes where the profile drops even faster \citep[cf.][]{Oh95,Penarrubia09}, causing the fitting routine to provide high values of $\gamma$ when trying to adjust the density profile. Leaving the outer profile index $\beta$ as a free parameter, i.e. using a $(\alpha,\beta,\gamma) = (1,\beta,1)$ model, we found indeed that on average the outer slope of SPH subhaloes is $\bar{\beta}=3.8$ while in the DM only case $\bar{\beta}=4.4$.
This is in agreement with the average values obtained for the Einasto shape parameter, which is higher in the SPH case with respect to the DM only scenario: the higher \nEin\, found in the SPH subhaloes indicates a cuspy inner profile, as expected if adiabatic contraction is acting on the central part of the structures, but also an outer density profile less steep than in the DM case, as confirmed by the $(1,\beta,1)$ model.\\
The outer profile of subhaloes in the CLUES gas-dynamical run is shallower than in the pure dark matter case because of the influence of tidal stripping whose effects, being present in both runs, are stronger in the DM case. Tidal stripping, which mainly acts on the outer part of the density profile, is able to remove more mass from a pure dark matter subhalo than an SPH one, owing to the deepening of the potential in the latter case, as shown in greater detail by \citet{Libeskind10}.

Finally, we also used the exponentially truncated profile, introduced by \citet{Kazantzidis04} to deal with the divergence of the cumulative mass distribution of haloes as $r \rightarrow \infty$, but we did not obtain improvements over the Einasto or P\&S profiles.
 
To further highlight the quality of the different models, in \Fig{fig:residuals} we present the residuals between the fits and the data for each subhalo in the SPH simulation as a function of $\rm log (r/r_{-2})$ (the plots look akin for the DM simulation and hence are omitted). Note that for the $(1,3,\gamma)$ model the point where the logarithmic slope of the density profile equals $-2$ occurs at a radius $r_{-2} = (2-\gamma) r_s$ for $\gamma <2$: thus, for a NFW profile $r_{-2}=r_s$ and for a M99 profile $r_{-2}=r_s/2$.

Neither the NFW, M99, or $(1,3,\gamma)$ profiles are well fitted over the whole radial range: while the $(1,3,\gamma)$ with a steep central slope may describe the data in the inner regions, it fails so in the outer parts.
On the contrary, the radial dependence of the residuals in the Einasto model is clearly minimized with respects to all the other models, being consistent with zero at every radial bin.
The case could be made that the $(1,3,\gamma)$ model performs as well as the Einasto model in the inner region of the density profile (which is also the region of interest with respect to the observations of the MW's dSphs). In order to assess the degree to which these results are affected by the choice of the radial range, we calculate the residuals, for every model, for only the innermost bins with $r<r_{-2}$, and found that the Einasto is still the best profile overall. 

We conclude that the Einasto model outperforms all the other proposed profiles in terms of quality of fit, giving, over the full radial range, an average value of $\overline{\Delta^2}=0.011$ in the SPH run and $\overline{\Delta^2}=0.012$ in the DM only run. We note that the the P\&S model also provides good results, though not as good as the Einasto model.

While it is somehow obvious that these 3-parameter models perform better than the 2-parameter ones (such as the NFW or M99 model), we are reassured by the fact that even after fixing the shape parameter \nEin\ (\nPS) of the Einasto (P\&S) profile to its mean value, therefore reducing the free parameters to two, we still obtain a mean goodness-of-fit which is lower than any other 2-parameter model (the Einasto profile, for example, provides $\overline{\Delta^2}=0.025$ for the SPH run and $\overline{\Delta^2}=0.028$ for the DM run).
Thus, the better performance of the Einasto profile is not just an artifact of having one free parameter more. Furthermore, our results are in agreement with those of other workers in the field \cite[e.g.][]{Springel08}.

In light of this we state that the need of a shape parameter $n$ to fully specify the mass profile of simulated DM and SPH subhaloes is an indication of the non-universality of their density profiles, as will be highlighted in the next section.

%%%%%%%%%%%%%%%%%%%%%%%%%%%%%%%%%%%%%%%%%%%%%%%%%%%%%
\section{Shape parameter -- subhalo mass relation}\label{sec:n_dependence}
%%%%%%%%%%%%%%%%%%%%%%%%%%%%%%%%%%%%%%%%%%%%%%%%%%%%%
In \Tab{tab:profiles} the average values of the Einasto shape parameter \nEin\ are shown. However, this shape parameter varies from subhalo to subhalo, spanning quite a large range  $0.4\lesssim$ \nEin  $\lesssim10.4$ in the SPH run (with a similar spread in the DM model). This naturally raises the question of whether this variation follows some rule or is random.

In \Fig{fig:nEin_mass}, the fitted Einasto shape parameter \nEin\, is plotted against subhalo mass for both the DM only run (left) and the SPH run (right). A clear correlation is immediately visible. This result forms one of the main findings of this paper: \textit{the Einasto shape parameter correlates directly with subhalo mass}. To quantify the \nEin-mass correlation, the Spearman rank coefficient\footnote{The Spearman rank coefficient is a non-parametric measure of correlation that assesses how well an arbitrary monotonic function describes the relationship between two variables, without making any other assumptions about the particular nature of the relationship between the variables. The closer the coefficient is to 1 the stronger the correlation between the two variables. We use the IDL routine R CORRELATE() to calculate it.} $S_r$ \citep{Kendall90} is calculated, yielding $S_r=0.70$ for the SPH run, with a significance of practically zero confirming a strong correlation: the most massive objects have a higher value of the shape parameter, while less massive ones have smaller values. In other words, low mass substructures are well fit by a inner density profile shallower than a NFW one, with a steep outer slope, while the higher mass objects are fit by a steep, cuspy-like inner profile. Convergence studies (wherein the number of radial bins used for profile fitting is drastically increased) have been performed in order to ascertain the applicability of an Einasto profile to our subhaloes. These tests have revealed that our fitting procedure is robust and not a result of the sampling. Similar results and $S_r$ values are found for the DM only run. One might argue that the subhalo's mass may be seen as a rather ill-defined quantity, and a better proxy for mass \citep[e.g.][]{Knebe11b} should be the peak of the rotation curve $V_{\rm max}$: when replacing the mass on the $x$-axis of \Fig{fig:nEin_mass} with $V_{\rm max}$ we actually do not find any substantial change in the correlation, strongly confirming it.

Since Fig.~\ref{fig:nEin_mass} shows the same \nEin-mass relation for both the DM only and the SPH run, we conclude that the mechanism responsible for this relation must be dynamical and hence is likely to be tidal stripping. To determine the influence of tidal stripping, the properties of subhaloes at infall time, defined as the last time a subhalo crossed a sphere of physical radius $300$kpc from the host's center \footnote{Using the first infall time provides similar results.}, have been examined.

\begin{figure}\begin{center}
  \includegraphics[width=3.3in]{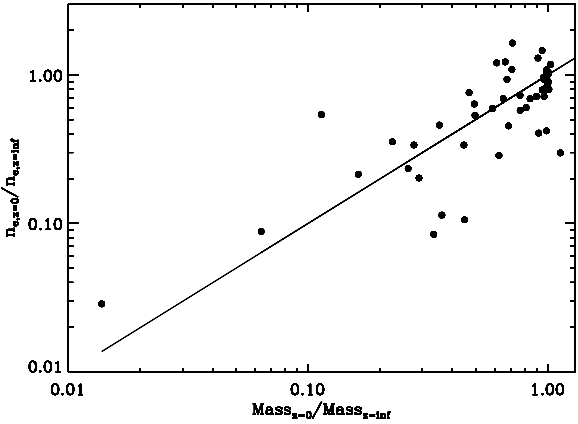}
\caption{Reduction of the shape parameter \nEin\ as a function of the mass loss between $z=0$ and $z_{\rm infall}$ for the SPH subhaloes. A similar behavior is found in the DM only run, thus not shown here. The best fit curve, which has unitary slope, is shown as a solid line.}
\label{fig:nEin_infall}
\end{center}
\end{figure}

At infall time the subhaloes' density is well described by both an Einasto and a NFW profile, as expected for field haloes. Using an Einasto profile, we find that the more mass lost since $z_{\rm infall}$, the lower the value of \nEin\ at $z=0$, as presented in \Fig{fig:nEin_infall} for the SPH case. There is thus an evident correlation between the amount of stripped material and the reduction of \nEin\ for each subhalo. Again, this relation is quantified by the Spearman rank coefficient, $S_r=0.68$, and showing the best curve fit, which has a unitary angular coefficient, as a solid line in \Fig{fig:nEin_infall}. 
A similar dependence is found in the DM only run, thus not shown here, corroborating our findings that tidal stripping is the main mechanism able to modify the density profile of subhaloes. 

Many authors \citep[e.g.][]{Kazantzidis04,Springel08,Hayashi03,Penarrubia10} have shown that tidal stripping acts mainly to modify the outer region of a profile. Since a steepening of the outer profile entails a reduction of \nEin\, exactly as observed in our simulations, we therefore conclude that the lowering of the subhaloes' shape parameter between $z_{\rm infall}$ and $z=0$ is primarily due to stripping effects. This finding is indeed in agreement with the recent work of \citet{Vera-Ciro12} who show that very heavily stripped objects have on average smaller \nEin, because of a steepening of the outer density profile. It is also worth noting that tidal stripping has been shown by \cite{Hayashi03} to not only affect the subhaloes' outer regions: as a substructure loses mass, the central density will also decrease significantly (although the slope of the inner density profile remains unchanged).

A word of caution is necessary at this point. The Einasto model's shape parameter \nEin\, describes simultaneously the slope of the inner and outer profile, in a single number. This can be both an advantage and a disadvantage. Indeed for some subhaloes in our simulations the value of the scale radius $r_{-2}$ is close to the the innermost converged radius while the outer profile is resolved with many radial bins. In these cases, if the outer profile steepens yet the structure of the inner part remains unchanged, the Einasto fit would return a lower value of \nEin\, due to the fitted profile being dominated by the steepened outer part. This is the case, for example, of heavily stripped objects $(M_{z=0}/M_{z_{infall}}\sim20\%)$ in the DM only run, in which the inner cuspy slope of the substructure is retained after infall, while the outer profile has been steepened by tidal stripping, thus providing a small fitted \nEin\ value.
Care must be taken not to interpret these cases as becoming cored, in the inner region, due to tidal stripping. That said even in these cases, Einasto models provide accurate fits to the density profile -- more accurate than any of the other models as could be seen in \Fig{fig:residuals} and \Tab{tab:profiles}.

While tidal stripping is the only relevant mechanism in the DM only run, other effects such as baryonic feedback have to be taken into account in the SPH case, since they can also contribute to changes in the density profile of subhaloes \citep{Zolotov12,Brooks12}. Indeed, as opposed to the DM only simulation, in the SPH run we observe in some subhaloes a change to their inner structure \citep[cf.][]{DiCintio11}. Since, as mentioned above, tidal stripping does not change the inner slope of substructures, the only mechanisms able to alter the density at small radii must have a baryonic origin.
\citet{DiCintio11} showed that the inclusion of baryons has a twofold effect, increasing or decreasing its central density according to an adiabatic contraction \citep{Blumenthal86} or outflows \citep{Navarro96b, Governato12} model.

To shed more light onto the effects of baryons, a one to one comparison of the density profiles of those subhaloes that can be cross-identified in the DM and SPH run has been performed \cite[as in][]{DiCintio11}. The subhaloes which experienced an expansion in the SPH run, with respect to their DM only partner, have lost all their gas at redshift zero, and they do not show any sign of star formation between $z_{\rm infall}$ and $z=0$. These subhaloes have an inner density profile shallower than the corresponding sister DM subhalo. On the other hand, those objects which have undergone adiabatic contraction in the SPH run \citep[cf.][]{DiCintio11}  still retain some gas at $z=0$ and their star formation appears to be on going even after infall. These subhaloes have a high value of the shape parameter \nEin, which is now well describing a steeper inner density profile caused by adiabatic contraction. In any case, the fits are still dominated by the outer profile, steepened by tidal stripping, where most of the bins lie and hence higher resolution simulations are needed to verify the effective creation of a core in objects with small \nEin\ values: the interplay of these two contrasting effects, i.e. outflows vs adiabatic contraction, and a deep analysis of the repercussions on the inner density profile of substructures in cosmological simulations will be explored in detail in a companion paper (Di Cintio et al., in prep.).

Some other important conclusions can now be drawn from \Fig{fig:nEin_mass}. Firstly, there is no evidence for any universal profile in simulated substructures. Secondly, we note that the majority of the subhaloes in both runs tend to have a small \nEin, while only the most massive ones (mostly the adiabatically contracted ones) have a high \nEin, as large as \nEin=10.4 in the SPH run and \nEin=9.4 in the DM only one. This finding, as well as the goodness of the Einasto profile, has been confirmed from an observational point of view by the recent work of \citet{DelPopolo12}, who used high quality rotation curves data of dwarf galaxies to show that the preferred fitting function is given exactly by the Einasto model and that the majority of the dwarfs tend to have shallow profiles (their Fig.~3). Our mean shape parameter in the hydrodynamical run, $\overline{n}_{\rm E}=4.8$, as well as their Einasto mean shape parameter, $\overline{n}_{\rm E,D.P.}=3.05$, are both lower than the corresponding \nEin\ for dwarf size objects found in previous dark matter simulations \citep{Merritt05,Navarro04}. In that regards we need to mentioned that previous results coming from such dark matter only simulations, where cluster- and galaxy-sized haloes have been studied, showed instead a decreasing of \nEin\ for increasing halo mass \citep{Navarro04,Merritt05,Graham06III,Prada06,Gao08,Hayashi08,Navarro10}. Our study indicates that there is a turnover, such that the trend with mass is reversed for low mass galaxies (at least satellites), with both SPH and DM simulations having a positive correlation of \nEin\ with mass. The main difference between previous studies and this work is, besides the less massive objects considered here, the fact that our objects are subhaloes, and thus obviously affected by tidal stripping, as mentioned above. Finally it must be noticed that the range of variation of the \nEin\ shape parameter found in our simulation is very large, spanning the interval  $0.4\lesssim$ \nEin  $\lesssim10.4$ in the SPH case: remarkably the same large range has also been found in \citet{DelPopolo12}, with $0.29<n_{\rm E,D.P.}<9.1$, as well as in the recent work of \citet{Vera-Ciro12}, based on semi-analytical models of galaxy formation.
Furthermore, in the observational paper of \citet{Chemin11}, the authors used the Einasto model to fit the rotation curves of the THINGS\footnote{http://www.mpia.de/THINGS/Overview.html} galaxies and show that the shape parameter is near unity on average for intermediate and low mass halos, while it increases for higher mass haloes, being correlated with the halo virial mass as we find in this work.

%%%%%%%%%%%%%%%%%%%%%%%%%%%%%%%%%%%%%%%%%%%%%%%%%%%%%
\section{New observational constraints for the satellite galaxies of the Milky Way}\label{sec:observation}
%%%%%%%%%%%%%%%%%%%%%%%%%%%%%%%%%%%%%%%%%%%%%%%%%%%%%
We now move to a practical application of our findings, only focusing on the properties of SPH subhaloes, which are obviously closer to reality than their DM only counterparts. 
In \citet{Boylan11} the observational constraints, used to establish if subhaloes found in cosmological simulations are possible hosts of the known Milky Way dwarf spheroidals, were based on the assumption that the underlying dark matter halo of these dSphs follows a NFW profile. Given our findings, however, it is clear that since the Einasto model provides the best fit to the density profile of both DM and SPH subhaloes, those observational constraints have to be modified.\footnote{\citet{Vera-Ciro12} reached similar conclusions, using semi-analytical galaxy formation models, while this manuscript was being prepared for submission.} Note that in Eq.5-7, for clarity, we will omit the subscript $_E$ from the Einasto shape parameter, simply referring to it as $n$.
The circular velocity of an Einasto profile follows
\beq
v^2(r)\propto \gamma(3n,x),\label{velox}
\eeq

\noindent
where 
\beq
\gamma(3n,x)=\int_{0}^x e^{-t}t^{3n-1}dt
\eeq

\noindent
is the lower incomplete gamma function and $x =2n(r/r_{-2})^{1/n}$. To find the radius \Rmax\ at which $dv(r)/dr=0$ we numerically solve,

\beq
\gamma(3n,2n(\frac{R_{\rm max}}{r_{-2}})^{1/n})  = 2^{3n}n^{3n-1} (\frac{R_{\rm max}}{r_{-2}})^3 e^{-2n(\frac{R_{\rm max}}{r_{-2}})^{1/n}}    \ .
\eeq

\noindent
The relation between \Rmax\ and $r_{-2}$, that we need in order to compute the observational constraints, varies depending from the value of the shape parameter \nEin\ \citep[see, for example, Fig.~2 in][]{Graham06}. Given the fact that the mass density profile of the faint dSphs is still uncertain \citep[e.g.][]{Walker11,Wolf12} we prefer to use the conservative limits given by the highest and the lowest values of the shape parameter \nEin\ as obtained from our hydrodynamical simulation. In the SPH run the smallest \nEin$=0.4$ corresponds to a relation $R_{\rm max}=1.447 r_{-2}$, while the largest \nEin$=10.4$ gives  $R_{\rm max}=2.348 r_{-2}$.
Using these constraints, i.e. the assumption of an Einasto model and the corresponding range of \nEin -values, we computed the curves in the \Vmax-\Rmax\ plane for the nine brightest classical dSphs of the MW, namely CvnI, Carina, Draco, Fornax, Leo I, Leo II, Sextans, Sculptor and Ursa Minor, which all have $M_V\lesssim-8.8$ (we excluded Sagittarius as in \citet{Boylan11} since it is far from dynamical equilibrium): these curves are constructed by normalizing each dwarf to its observationally derived values of half-light mass, $M_{1/2}$, and radius, $r_{1/2}$, from \citet{Wolf10}, who showed that any uncertainty on the stellar velocity dispersion anisotropy is minimized at this radius, leading to accurate estimation of $v(r_{1/2})$. 

\begin{figure*}\begin{center}
$\begin{array}{cc}
\includegraphics[width=3in]{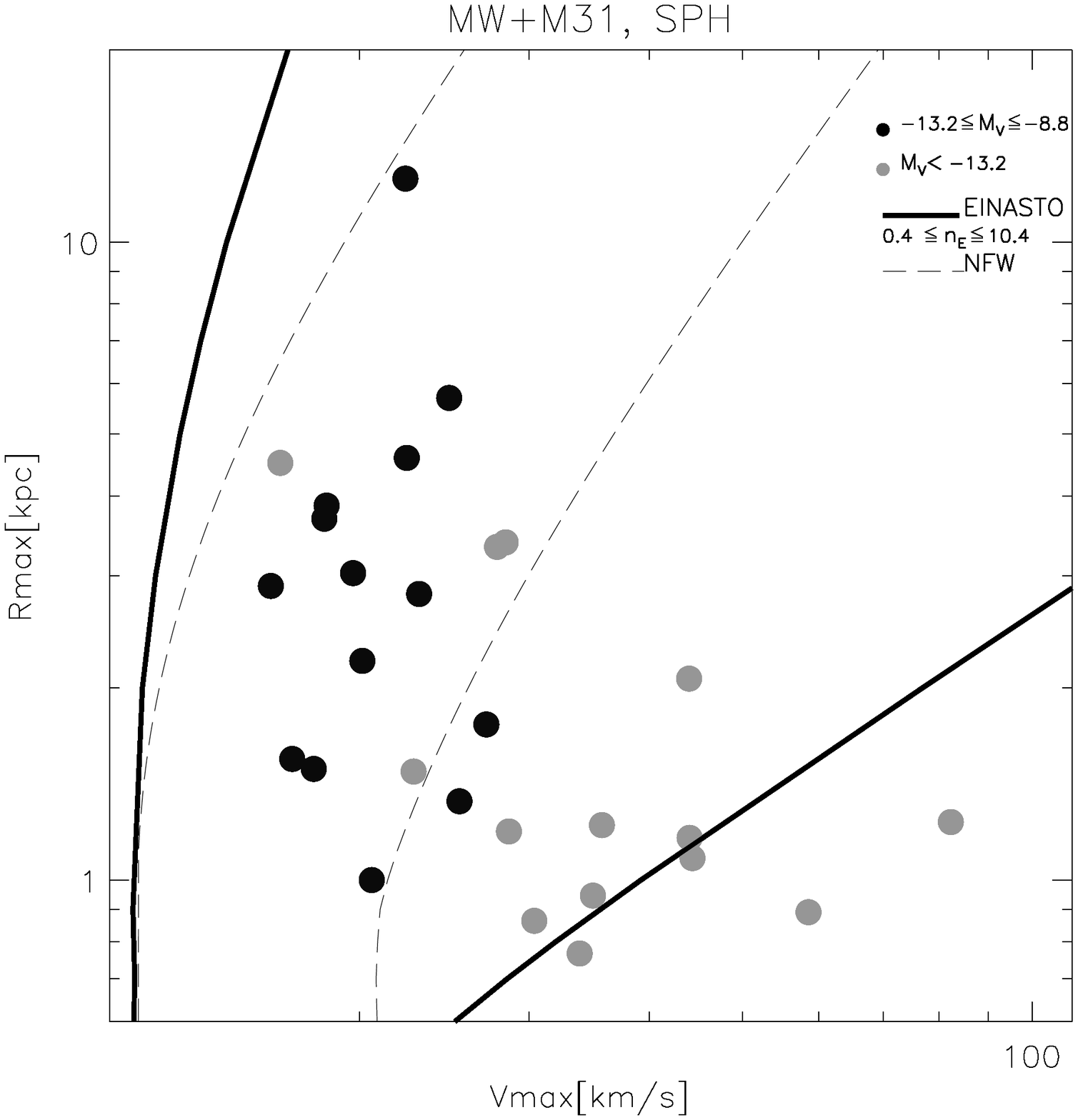}
\includegraphics[width=3in]{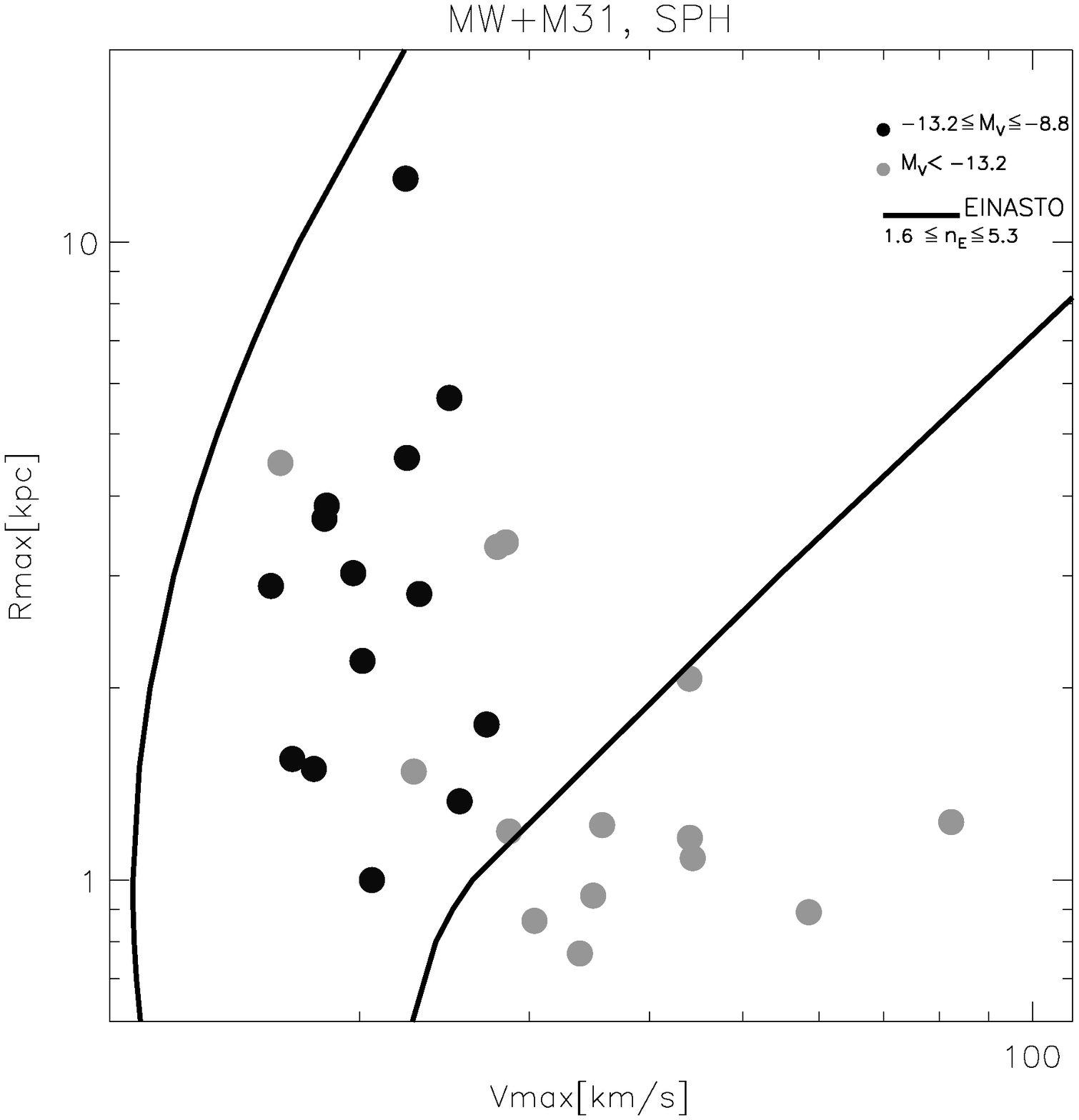}
\end{array}$
 \caption{Left panel, \Vmax-\Rmax\ pairs for the SPH subhaloes within the MW and M31 hosts. The subhaloes have been color-coded by their luminosity: in black the ones corresponding to the luminosity of the observed classical dSphs, in grey the ones which are brighter than $M_V=-13.2$. The $2\sigma$ observational constraints for the MW's dSphs are indicated as a solid line for the Einasto profile with shape parameter $0.4\lesssim n_{\rm E}\lesssim 10.4$, obtained considering all the subhaloes with $M_V\lesssim -8.8$, and as dashed line for the NFW profile. Right panel, same criterion to color-code the subhaloes, although this time the $2\sigma$ observational constraints for the MW's dSphs are derived using an Einasto profile with shape parameter $1.6 \lesssim n_{\rm E}\lesssim 5.3$, obtained considering only the subhaloes with $-13.2\lesssim M_V\lesssim -8.8$.}
\label{fig:several_profiles}
\end{center}
\end{figure*}

\begin{figure}
\includegraphics[width=3.3in]{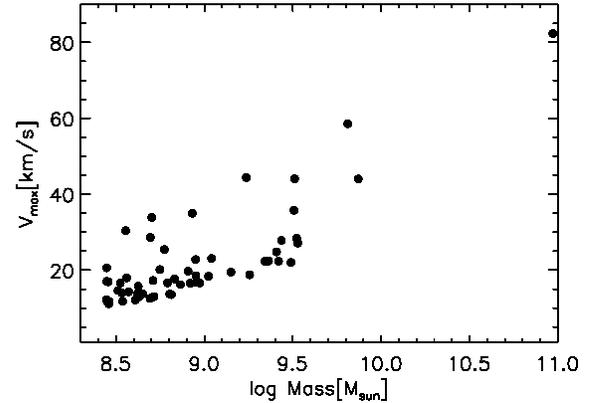}
\caption{Correlation of the subhaloes maximum circular velocity with the subhaloes' masses in the SPH simulation.}
\label{fig:vmax_mass}
\end{figure}

In the left panel of \Fig{fig:several_profiles} we show the maximum circular velocity $V_{\rm max}$ and its corresponding radius $R_{\rm max}$ for all the SPH subhaloes, within the MW and M31 hosts, whose luminosity is at least as high as the Draco's one, i.e. $M_V\lesssim -8.8$. 
With respect to \Fig{fig:nEin_mass} we excluded here objects with a luminosity lower than Draco, but verified that the interval for the shape parameter \nEin\ is still the same.
In \Fig{fig:vmax_mass} we also show the numerically derived \Vmax-mass relation for the SPH subhaloes, which is useful to derive the range of masses associated to a specific value of \Vmax\ . 
The grey symbols in \Fig{fig:several_profiles} correspond to the subhaloes that are brighter than Fornax, which is the brightest classical dwarf considered here to construct the observational constraints having $M_V=-13.2$, and the black circles indicate all the remaining subhaloes with luminosity $-13.2\lesssim M_V\lesssim -8.8$. We also plot the newly constrained observational limits, as solid lines, coming from the assumption that the MW's dSphs are embedded in haloes that follow the Einasto profile with varying shape parameter \nEin\ between $0.4\lesssim n_{\rm E} \lesssim10.4$, and, as dashed lines, the previously used constraints coming from the NFW model.

We observe that, while the employment of an Einasto profile leads to a good agreement between observations and the SPH subhaloes,\footnote{We find a complete agreement between observations and DM only subhaloes, which are though not shown in the plot for clarity.} it still appears to be not sufficient to explain the $R_{\rm max}$-$V_{\rm max}$ pairs of the most massive SPH subhaloes which still lie in the lower right part of the plane, outside the constraints. Note that with a shape parameter varying between $0.4\lesssim n_{\rm E} \lesssim10.4$ we have allowed the observational constraints to cover a wider range,  in the \Rmax-\Vmax\ plane, with respect to the NFW constraints, but even this assumption is not enough to reconcile simulation and observation. However, those massive SPH objects, which we color-coded in grey, appear to have a luminosity, $M_V\lesssim-13.2$, not compatible with any of the satellites used to derive the observational constraints. 

We remind the reader that the data plotted in \Fig{fig:several_profiles} refer to both the MW and M31 galaxies: a total number of 14 subhaloes brighter than Fornax is thus found within the two hosts (and only 11 if we relax the magnitude cut from $M_V<-13.2$ to $M_V<-14$).
For each host halo we therefore have 5 to 7 objects brighter than the classical dwarfs used to compute the observational constraints. Three of them may be associated with the Large and Small Magellanic Clouds (LMC and SMC, respectively) and the Sagittarius galaxy. In fact, objects with $M_V \lesssim -16.2$ can be conservatively considered as the analogous of the LMC and SMC: we should therefore exclude these simulated subhaloes from the discussion. 

With these associations there are, within each host, only two to four subhaloes left with $-16.2 \lesssim M_V \lesssim -13.2$ which do not have a counterpart in the real universe: such a small sample, i.e. the $\sim10\%$ over the total number of objects found within each halo in the SPH run, can be explained as a statistical fluctuation due to our small number statistics. 
In the future, to confirm this halo-to-halo variation in the subhaloes population, it will be necessary to study many realization of a high resolution MW-like object.  
In this work the luminosity function, averaged over the MW and M31 subhaloes, has been shown to be in agreement with the observational data of Milky Way like galaxies \citep{Strigari12}, while slightly deviating from the Milky Way itself in the interval $-16\lesssim M_V\lesssim -13$ \citep{Knebe11a} exactly because of these two to four overabundant objects in this range.
Moreover, we remind that our simulation also reproduces the luminosity vs velocity dispersion correlation observed for the satellite galaxies of MW and M31 \citep{Walker09}, as shown in \citet{Knebe11a}.

As an additional remark, we note that most of the brightest subhaloes in our simulation are the ones that experienced adiabatic contraction, being situated in the lower right part of the \Vmax-\Rmax\ plane at redshift $z=0$, as studied and explained in \citet{DiCintio11}. These subhaloes are substantially different from the ones found in the work of \citet{Vera-Ciro12}, who used dark matter only simulations with semi-analytical galaxy formation models that do not show adiabatic contraction: their most luminous, brightest objects are found in the upper-right of the \Vmax-\Rmax\ plane, contrary to what we obtained in our hydrodynamical simulations.

We now proceed to again compute the shape parameter range based only upon those subhaloes that satisfy the luminosity requirement, i.e. those objects whose $M_V$ is within the range of the observed dSphs luminosity. We were therefore able to restrict the range, finding a shape parameter lying within $1.6\lesssim n_{\rm E} \lesssim 5.3$, with a mean value $\overline{n}_{\rm E} =3.2$.
In the right panel of  \Fig{fig:several_profiles} we use the same black-grey colouring scheme for the SPH subhaloes as before, and we plot the observational limits based on the newly constrained range for the shape parameter $1.6\lesssim n_{\rm E} \lesssim 5.3$.
The result is a perfect agreement between the expected \Vmax-\Rmax\ values of the observed dSphs and the  \Vmax-\Rmax\ pairs of the simulated subhaloes with corresponding luminosities. 

We conclude that our findings, based upon self-consistent hydrodynamical simulations of a constrained Local Group in a cosmological context, strongly supports the notion that the observed satellite galaxies of the Milky Way are actually compatible with being embedded in dark matter haloes whose density profiles show considerable differences, following an Einasto model with $1.6\lesssim n_{\rm E}\lesssim 5.3$ and mean value $\overline{n}_{\rm E} =3.2$: the majority of the dSphs may have an inner profile shallower than the previously assumed NFW one while an outer profile steepened by tides.

%%%%%%%%%%%%%%%%%%%%%%%%%%%%%%%%%%%%%%%%%%%%%%%%%%%%%
\section{Conclusion and discussion}\label{sec:conclusion}
%%%%%%%%%%%%%%%%%%%%%%%%%%%%%%%%%%%%%%%%%%%%%%%%%%%%%
Using a constrained simulation of the Local Group, performed within the CLUES project, it has been shown that:

\begin{itemize}
 \item the density profile of subhaloes in both dark matter only and hydrodynamical simulations is best approximated by an Einasto profile in which the shape parameter \nEin\ is free to vary, and that
  \item there is a clear trend of growing \nEin\ with increasing subhaloes mass, in both the dark matter only and hydrodynamical run.
\end{itemize}

\noindent
The structural effect associated with tidal stripping is likely the main mechanism able to modify the subhaloes' density profile: the effect of mass loss due to tidal stripping is the reduction of the shape parameter \nEin\ between the infall and the present time. A correlation between \nEin\ and the amount of stripped material has also been argued in \citet{Vera-Ciro12}. Differences in the inner profile of subhaloes, between the pure DM and SPH run, can instead be attributed to baryonic processes \citep{DiCintio11}: they result in adiabatic contraction of the dark matter halo when gas is retained in the central regions and star formation is still on going after infall, and in expansion when gas is removed from the dwarfs due to stellar feedback and ram pressure stripping, with no signs of star formation after the subhalo has entered the host's halo.
These baryonic effects, acting mainly on the inner part of the density profile, do not modify the overall \nEin\ -mass relation which is driven by tidal stripping also in the SPH case.

The majority of our SPH subhaloes have a small \nEin, as reported in the right panel of \Fig{fig:nEin_mass}: remarkably, \citet{DelPopolo12} found similarly small values of \nEin\ in observed dwarf galaxies, using high quality rotation curves. Moreover, evidences of the fact that at least some of the MW's dSphs may have a shallow profile, compatible with a small \nEin, are given in \citet{Walker11}, who showed that the profiles of the Fornax and Sculptor dSphs are consistent with cores of constant density at a high confidence level. Nevertheless, the actual mass profile of the MW's dSphs is still uncertain: \citet{Wolf12} claimed that, even with an isotropic velocity dispersion, not all the dSphs prefer constant-density cores and that, instead, some of them favor a cuspy inner profile. All these findings do no longer support the notion of a universal subhalo mass profile; subhaloes of differing mass cannot be rescaled to have self-similar profiles: their mass (or so to speak size) matters.

In light of these results we revisited the \citet{Boylan11} observational limits for possible hosts of the MW's dSphs, assuming that the latter are embedded in haloes that follow an Einasto profile, as opposed to the earlier assumption of NFW profile, with variable shape parameter \nEin, and using the conservative limits $0.4\lesssim n_{\rm E} \lesssim 10.4$ provided by our hydrodynamical simulations. 
While using the Einasto profile is enough to completely explain the maximum velocity of the most massive DM subhaloes, an issue still remains with respect to the most massive SPH subhaloes: these objects experienced adiabatic contraction \citep{DiCintio11} and their \Rmax-\Vmax\ pairs are still lying outside the expected observational constraints.  However, these subhaloes appear to be brighter than Fornax, which is the brightest dSph used when constructing the observational constraints: thus, they should not be considered in the comparison. Once the Large Magellanic Cloud, Small Magellanic Cloud and Sagittarius galaxy analogues are removed, we still have two to four unaccounted objects per halo, whose luminosity is higher than the luminosity of the classical dwarfs: we argue that, being only $10\%$ of the total set, they can be interpreted as a statistical deviation.
 
Leaving only the SPH subhaloes with $ -13.2\lesssim M_V \lesssim -8.8$, i.e. those in agreement with the luminosity of the nine classical dSphs, we show that an Einasto profile with shape parameter $1.6 \lesssim n_{\rm E}\lesssim 5.3$ provides an accurate matching between simulations and observations, alleviating the "massive failures" problem first addressed in \citet{Boylan11}.  The mean value of the shape parameter for them is $\overline{n}_{\rm E} =3.2$, indicating that the majority of the MW's satellite galaxies are consistent with dark matter haloes whose profile is an Einasto one, steepened outside by the effects of tidal stripping and possibly shallower than the previously accepted NFW towards the center.

We further note that our simulated host haloes masses are at the low end of current observational estimates, and this may be one of the reason for having only a few objects in the luminosity range $-16\lesssim M_V\lesssim -13$: \citet{DiCintio11} suggested that the host halo mass is directly connected to the number of massive subhaloes found in simulations, when comparing the results of dark matter only simulations based on a WMAP3 versus WMAP5 cosmology, the latter showing a higher host halo mass and consequently a higher number of massive subhaloes. The dependence of the number of "too massive subhaloes" on halo mass has been further explored and quantified by \citet{Wang12}. 
\citet{Boylan11} extensively discussed the possibility that the reduction of the Milky Way mass could solve the problem, and recently \citet{Vera-Ciro12} suggested that a Milky Way mass $\sim 8\cdot10^{11}M_{\odot}$ provides a good match between observations and semi-analytical galaxy formation models.  In our simulations we have slightly lower masses for the MW and M31, between $5.5$ and $7.5\cdot10^{11}M_{\odot}$, these values being at the low end of mass estimates obtained using different methods \citep{Kara06,Watkins10,Deason12}. 
According to the model of \citet{Wang12}, a MW mass of $M_{MW}=5.5\cdot10^{11}M_{\odot}$ will give a $84\%$ probability of finding only three satellite galaxies with a circular velocity peak higher than \Vmax\ $>30$ km/s: this is what we expect since, apart from the LMC, SMC and Sagittarius, all the other classical dwarfs have been shown \citep[e.g.][]{Strigari10} to inhabit haloes with a maximum circular velocity below $30$ km/s. However, such a low mass for the Milky Way will reduce the probability that it hosts two satellites as the LMC and SMC; that our galaxy system is rare, with only $\sim3.5\%$ of the MW-like candidates having two satellites as bright as the Magellanic Clouds, has been found observationally by studies using the Sloan Digital Sky Survey \citep{Liu11,Guo11,Lares11,Tollerud11b}. We remark that, despite a lower Milky Way mass, subhalo density profiles should nevertheless be described by an Einasto model in order to properly match the kinematic of the observed classical dSphs with the subhaloes in hydrodynamical simulations.
Finally, even assuming a small mass for the Milky Way and an Einasto profile for its satellite galaxies, there is still a problem in assigning the correct halo masses to dwarf galaxies, as highlighted in \citet{Ferrero11}. These authors showed that the MW's dSphs, as well as many isolated dwarf galaxies with spatially resolved rotation curves and stellar mass $10^6 < M_{gal}/M_{\odot} < 10^7$, seem to live in haloes with $M<10^{10}M_{\odot}$, which is at odds with the abundance-matching prediction of \citet{Guo10} and \citet{Moster10}: the validity of \LCDM\ at such scales is still disputable.

%%%%%%%%%%%%%%%%%%%%%%%%%%%%%%%%%%%%%%%%%%%%%%%%%%%
\section*{Acknowledgements}
%%%%%%%%%%%%%%%%%%%%%%%%%%%%%%%%%%%%%%%%%%%%%%%%%%%
We kindly acknowledge Mia S. Bovill, James S. Bullock and Carlos A. Vera-Ciro for the fruitful discussions that lead to an improved version of the manuscript. We further thank the referee for the critical and helpful comments on the paper. 
The simulations were performed and  analyzed at  the Leibniz Rechenzentrum Munich (LRZ) and at the Barcelona Supercomputing Center (BSC). We thank DEISA for giving us access to computing resources in these centers through the DECI projects SIMU-LU and SIMUGAL-LU.
AK is supported by the {\it Spanish Ministerio de Ciencia e Innovaci\'on} (MICINN) in Spain through the Ramon y Cajal programme as well as the grants AYA 2009-13875-C03-02, AYA2009-12792-C03-03, CSD2009-00064, and CAM S2009/ESP-1496. YH has been partially supported by the Israel Science Foundation (13/08). NIL is supported by a grant by the Deutsche Forschungs Gemeinschaft.  GY acknowledges support also from MICINN under research grants AYA2009-13875-C03-02, FPA2009-08958 and Consolider Ingenio SyeC CSD2007-0050.

\bibliographystyle{mn2e}
\bibliography{archive}

\bsp

\label{lastpage}

\end{document}